\documentstyle[12pt]{article}
\newcommand{\doublespacing}{\let\CS=\@currsize\renewcommand{\baselinesstrech}
{2.0}\tiny\CS}

\begin{document}

\title{Phase properties of a new nonlinear coherent state}
\author{B.Roy.\thanks{E-mail : barnana@isical.ac.in} and
P. Roy\thanks{E-mail : pinaki@isical.ac.in}\\ Physics \& Applied
Mathematics Unit\\ Indian Statistical Institute \\ Calcutta  700035\\ India} 


\maketitle

\vspace*{1.5cm}

\centerline{\bf Abstract}

\vspace{0.3cm}

\thispagestyle{empty}

\setlength{\baselineskip}{18.5pt}

We study phase properties of a displacement operator type nonlinear coherent
state. In particular we evaluate the Pegg-Barnett phase distribution and
compare it with phase distributions associated with the Husimi Q function and the
Wigner function. We also examine number-phase squeezing of this state.

\section{Introduction}
Coherent states are important in different branches of physics particularly
in quantum optics. Historically coherent state of the harmonic oscillator
(which is the coherent state corresponding to the Heisenberg-Weyl algebra)
was first constructed by Schr\"{o}dinger \cite{Sch}. Subsequently coherent
states corresponding to various Lie algebras like Su(1,1), Su(2) etc. have also
been constructed and has been shown to play important roles in the
description of various quantum optical processes \cite{Pere,Zhang}.

Recently another type of coherent states called the nonlinear coherent states
\cite{filho} or the f-coherent states \cite{manko} have been constructed. In
contrast to the coherent states mentioned above these are coherent states
corresponding to nonlinear algebras. However nonlinear coherent states are not
mere mathematical objects. It has been shown \cite{filho} that
nonlinear coherent states are useful in connection with the motion of a
trapped ion. 

We note that unlike Lie algebras,the commutators of the generators of nonlinear
algebras is a nonlinear function of the generators. As a consequence it is
difficult to apply the BCH disentangling theorem \cite{loui} to construct 
displacement operator coherent states corresponding to nonlinear algebras.
To avoid this difficulty nonlinear coherent states were constructed as
eigenstates of a generalised annihilation operator \cite{filho,manko}. However,nonlinear
coherent states can still be constructed using a displacement operator,{\it{albeit}},
a modified one \cite{fern,bam} and it has been shown that such states exhibit
nonclassical behaviour \cite{roy00}. In the present paper we shall study
phase properties of such a nonlinear coherent state. In particular we shall
obtain the Pegg-Barnett phase distribution and compare
it with the phase distributions associated with Q-function and the Wigner
function. We shall also evaluate the number-phase uncertainty relation and 
examine number-phase squeezing of the nonlinear coherent state.
The organisation of the paper is as follows: in section 2 we derive phase
distributions and number-phase uncertainty relation for displacement operator
nonlinear coherent states; in section 3 we discuss numerical results obtained
using the results of section 2; finally section 4 is devoted to a conclusion.

\section{New nonlinear coherent states and their phase distributions}

To begin with we note that the generalised creation and annihilation operators
associated with nonlinear cohrent states are of the form \cite{filho,manko}
\begin{equation}
A^\dagger =  f(N)a^\dagger~~,~~A = a f(N)~~,~~N = a^\dagger a \label{comm}
\end{equation}
where $a^\dagger$ and $a$ are standard harmonic oscillator creation and annihilation
operators and $f(x)$ is a reasonably well behaved real function,called the 
nonlinearity function. From the relations (\ref{comm}) it follows that $A$,
$A^\dagger$ and $N$ satisfy the following closed algebraic relations:
\begin{equation}
[A,A^\dagger] = f^2(N)(N+1) - f^2(N-1)N~~,~~[N,A] = -A~~,~~[N,A^\dagger] = A^\dagger \label{deform}
\end{equation}
Thus (\ref{deform}) represent a deformed Heisenberg algebra whose nature of
deformation depends on the nonlinearity function $f(n)$. Clearly for $f(n)=1$
we regain the Heisenberg algebra. Nonlinear coherent states $|\chi>$ are
then defined as right eigenstates of the generalised annihilation operator
$A$ and in a number state basis is given by \cite{filho,manko}:
\begin{equation}
|\chi> = C \sum_{n=0}^\infty \frac{d_n}{\sqrt n!} \chi^n |n> \label{def1}
\end{equation} 
where $C$ is a normalisation constant and the coefficients $d_n$ are given by
\begin{equation}
d_0 = 1~~,~~d_n = [\Pi_{i=1}^n f(i)]^{-1}
\end{equation}

We note that the canonical conjugate of the generalised annihilation and
creation operator $A$ and $A^\dagger$ are given by \cite{roy00}
\begin{equation}
B = a \frac{1}{f(N)}~~,~~B^\dagger = \frac{1}{f(N)} a^\dagger
\end{equation}
Thus $A$ and $B^\dagger$ and their hermitian conjugates satisfy the algebras
\begin{equation}
[A,B^\dagger] = 1~~~,~~~[B,A^\dagger] = 1 \label{comm1}
\end{equation}
Now following ref \cite{roy00} we consider the operators 
$(\alpha=|\alpha|e^{i\phi})$
$$D_1(\alpha) = exp(\alpha B^\dagger - \alpha^* A)$$
\begin{equation}
D(\alpha) = exp(\alpha A^\dagger - \alpha^* B)
\end{equation}
and note that for two operators $X$ and $Y$ such that $[X,Y]=1$ the BCH disentangling 
theorem \cite{loui} is of the form
\begin{equation}
exp(\alpha X - \alpha^* Y) = exp(-\frac{\alpha \alpha^*}{2})exp(\alpha X)exp(-\alpha^* Y) \label{BCH}
\end{equation}
Then the nonlinear coherent state corresponding to the first of the two algebras in (\ref{comm1}) is
defined as $|\alpha>_1 = D_1(\alpha)|0>$. Now applying (\ref{BCH}) we obtain
\begin{equation}
|\alpha>_1 = c_1 \sum_{n=0}^\infty \frac{d_n}{\sqrt n!} \alpha^n|n>
\end{equation}
Comparing this with the nonlinear coherent state $|\chi>$(see (\ref{def1})) we
find that both are exactly the same(provided of course we use the same 
nonlinearity function in both the cases).

The new nonlinear coherent state is then defined as $|\alpha> = D(\alpha)|0>$ i.e,
it is the coherent state corresponding to the second algebra in (\ref{comm1}). As
before using the relation (\ref{BCH}) we obtain
\begin{equation}
|\alpha> = c \sum_{n=0}^\infty \frac{d_n^{-1}}{\sqrt n!}\alpha^n |n> \label{new}
\end{equation}
where $c$ is a normalisation constant which can be determined from the condition
$<\alpha|\alpha>=1$ and is given by
\begin{equation}
c^2 = [\sum_{n=0}^\infty \frac{d_n^{-2}}{n!}(\alpha^* \alpha)^n]^{-1} 
\end{equation}

We now consider the phase probability distributions for the new nonlinear
coherent state (\ref{new}). According to the Pegg-Barnett formalism \cite{pegg}
a complete set of $(s+1)$ orthonormal phase states $\theta_p$ are defined by
\begin{equation}
|\theta_p> = \frac{1}{\sqrt{(s+1)}} \sum_{n=0}^s exp(in\theta_p)|n>
\end{equation}
where $|n>$ are the number states which spans the $(s+1)$ dimensional state
space and $\theta_p$ are given by
\begin{equation}
\theta_p = \theta_0 + \frac{2\pi p}{s+1}~~,~~p = 0,1,2,...,s \label{theta}
\end{equation}
In (\ref{theta}) $\theta_0$ is arbitrary and indicates a particular basis in
the phase space. The hermitian phase operator is then defined as
\begin{equation}
\Phi_\theta = \sum_{p=0}^s \theta_p|\theta_p><\theta_p|
\end{equation}
The expectation value of the phase operator with respect to the new nonlinear
state $|\alpha>$ is given by
\begin{equation}
<\alpha|\Phi_\theta|\alpha> = \sum_{p=0}^s \theta_p|<\theta_p|\alpha>|^2 \label{limit}
\end{equation}
where $|<\theta_p|\alpha>|^2$ is the probability of being in the state $|\theta_p>$.
Then in the limit $s\rightarrow \infty$ we get from (\ref{limit})
\begin{equation}
<\alpha|\Phi_\theta|\alpha> = \int_{\theta_0}^{\theta_0+2\pi} \theta P(\theta)d\theta
\end{equation}
where the continuous probability distribution $P(\theta)$ is given by
\begin{equation}
P(\theta) = lim_{s\rightarrow \infty} \frac{s+1}{2\pi} |<\theta_p|\alpha>|^2 \label{P}
\end{equation}
Now choosing $\theta_0$ as
\begin{equation}
\theta_0 = \phi - \frac{\pi s}{s+1}
\end{equation}
and using (\ref{P}) we obtain the Pegg-Barnett phase probability distribution 
for the new nonlinear coherent states (\ref{new}):
\begin{equation}
P_{PB}(\theta) = \frac{1}{2\pi} \left[1 + 2c^2 \sum_{n>k} \frac{d_n^{-1}d_k^{-1}}
{\sqrt{n!k!}}cos{[(n-k)\theta]} \right]~~,~~-\pi \leq \theta \leq \pi \label{PB}
\end{equation}
With the phase probability distribution known various quantum mechanical averages
in the phase space can be obtained using this function. For example the
phase variance is given by
\begin{equation}
<(\Delta \Phi_\theta)^2> = \int_{-\pi}^\pi \theta^2 P(\theta) d\theta 
= \frac{\pi^2}{3}+4c^2 \sum_{n>k} \frac{d_n^{-1}d_k^{-1}}{\sqrt{n!k!}} 
\frac{(-1)^{n-k}}{(n-k)^2}
\end{equation}
It may be noted that since $N$ and $\Phi_\theta$ are canonically conjugate operators
they obey the uncertainty relation
\begin{equation}
<(\Delta N)^2><(\Delta \Phi_\theta)^2>~~\geq~~\frac{1}{4} |<[N,\Phi_\theta]>|^2 \label{uncer}
\end{equation}
where $<(\Delta X)^2> = <X^2> - <X>^2$ and the right hand side of (\ref{uncer})
is given by
\begin{equation}
[N,\Phi_\theta] = i[1-2\pi P(\theta_0)]
\end{equation}
Now to examine number-phase squeezing we introduce the following squeezing
parameters:
\begin{equation}
S_N = \frac{2<(\Delta N)^2>}{|<[N,\Phi_\theta]>|} - 1~~,~~S_\Phi = \frac{2<(\Delta \Phi_\theta)^2>}{|<[N,\Phi_\theta]>|} - 1 \label{squeezing}
\end{equation}
If $S_N < 0$ ($S_\Phi <0$) then the nonlinear coherent state is number(phase)
squeezed.

We note that the phase quasiprobability distributions $P_{Q,W}(\theta)$
associated with the Husimi Q-function and the Wigner function can be obtained
by integrating these functions over the radial variable $|\beta|$. The forms
of these distributions are given by
\begin{equation}
P_{Q,W}(\theta) = \frac{1}{2\pi} \left[1 + 2c^2 \sum_{n>k} 
\frac{d_n^{-1} d_k^{-1}}
{\sqrt{n!k!}} cos[(n-k)\theta]F(n,k) \right]~~,~~-\pi\leq \theta \leq \pi 
\label{QW}
\end{equation}
where the coefficients $F(n,k)$ in the case of Q-function are given by \cite{tanas1,tanas2}
\begin{equation}
F(n,k) = \frac{\Gamma(\frac{n+k}{2}+1)}{\sqrt{n!k!}}
\end{equation}
while in the case of Wigner function they are given by \cite{tanas2,garra}
$$F(n,k) = 2^{(n-k)/2}\sqrt{\frac{k!}{n!}}\frac{\Gamma(\frac{n}{2}+1)}{(\frac{k}{2})!}
~~,{\rm n}~~{\rm even}$$
\begin{equation}
 = 2^{(n-k)/2}\sqrt{\frac{k !}{n !}}\frac{\Gamma(\frac{n+1}{2})}
{(\frac{k-1}{2})!}~~,{\rm n}~~{\rm odd}
\end{equation}

\section{Phase properties}

We shall now analyse various phase distributions for the nonlinear coherent
state. However,before we do this it is necessary to specify a nonlinearity
function $f(n)$. It is clear that for different choices of the nonlinearity function
we shall get different nonlinear coherent states. In the present case we choose
a nonlinearity function which has been used in the description of the motion 
of a trapped ion \cite{filho}:
\begin{equation}
f(n) = L_n^1(\eta^2)[(n+1)L_n^0(\eta^2)]^{-1} \label{nonlin}
\end{equation}
where $\eta$ is known as the Lamb-Dicke parameter and $L_n^m(x)$ are 
generalised Lagurre polynomials. We shall now evaluate the distribution
functions (\ref{PB}) and (\ref{QW}) with the nonlinearity function given by
(\ref{nonlin}).

In fugure 1 we plot Pegg-Barnett phase distribution $P_{PB}(\theta)$ against
$\theta$ keeping $\alpha$ fixed and using different values of $\eta$ for
the three curves. From figure 1 we find that for lower values of
$\eta$ the distribution is broad at the top. However as $\eta$ increases peaks
begin to develop slowly and for a reasonably large value of $\eta$ there are two
well developed peaks at $\theta=\pm \pi/2$. The appearence of the peaks is an
indication of quantum interference.

In figure 1a we plot the Pegg-Barnett phase distribution keeping $\eta$ fixed
at $.8$ and varying $\alpha$. From the figures it is seen that the qualitative
features of the distribution remains essentially the same when one of the parameters
is kept fixed while the other varies. However it may be noted that as $\alpha$
increases the peak structure becomes more and more prominent. Interestingly
for $\alpha=.37$(it is the value where the phase variance is minimum, see below)
the distribution shows practically no bifurcation.

In figure 2 we plot the three distributions $P_{PB}(\theta)$ and $P_{Q,W}(\theta)$
for the same values of the parameters. From the figure we find that although
the distributions are of the same form they are not quite the same. It is 
seen that $P_{PB}$ is roughly intermediate between $P_Q$ and $P_W$. Also the
Wigner distribution $P_W$ has the sharpest peaks while the Husimi Q
distribution $P_Q$ is the broadest.
However in the present case $P_W$ does not assume any negative value.

In figure 3 we have plotted phase variance against $\alpha$ for $\eta = .8$.
From the figure we find that phase variance decreases upto a certain value of
$\alpha$ and then again starts increasing. From the figure we find that phase
variance assumes the minimum value at $\alpha = .37$. Thus for $\alpha = .37$
and around this value of $\alpha$ the best measurement of phase is possible.
We note that the parameter values are not special but the phase variance shows
the same pattern for other parameter values too. 

In figure 4 we plot the squeezing parameters $S_N$ and $S_\Phi$. From the figure
we find that $S_N < 0$ for a considerable range of $\alpha$. This implies that
the nonlinear coherent state exhibits squeezing in the N component. However,
$S_\Phi$ is always positive implying absence of squeezing in the $\Phi$ 
component. 

We note that it is interesting to compare the squeezing behaviour of the 
new coherent state (\ref{new}) and the one given by (\ref{def1}). In figure 5
we plot the graphs of $S_N$ corresponding to these states. From the figure it
is seen that while for low values of $\alpha$ the coherent state (\ref{def1})
is more squeezed than (\ref{new}),for larger values as well as a larger range
of $\alpha$,the coherent state (\ref{new}) remains squeezed while (\ref{def1})
does not remain so.

Now we compare the phase squeezing of the nonlinear coherent states (\ref{def1})
and (\ref{new}). From fig 6 we find that $S_\Phi < 0$ for (3) while $S_\Phi > 0$
for (10). Thus from (\ref{squeezing}) it follows that the nonlinear coherent
state (\ref{def1}) exhibits phase squeezing while (\ref{new}) does not.

Finally to examine the number-phase uncertainty relation (\ref{uncer}),we
consider the quantity $F(\alpha) = \sqrt{<(\Delta N)^2><(\Delta \Phi_\theta)^2>}-
\frac{1}{2} |<[N,\Phi_\theta]>|$. Thus $F(\alpha)\geq 0$ and $F(\alpha) = 0$ would 
imply that the state is an intelligent state. On the other hand a nonzero
value of $F(\alpha)$ is a measure which indicates how much the state is away from
being an intelligent state. In figure 7 we plot $F(\alpha)$ against $\alpha$
for $\eta = .8$. From the figure it is seen that $F(\alpha)$ is nonzero and
has an increasing trend. The maximum in fig 7 indicates how much the nonlinear
coherent state (\ref{new}) can be different from an intelligent state. 
Thus we conclude that the nonlinear coherent state (\ref{new})
is not an intelligent state.

\section{Conclusion}
In this article we have considered a class of nonlinear coherent states constructed
using an operator similar to the displacement operator. We have examined a 
number of their phase properties. In particular we have computed various
phase distributions and compared them. Also it has been shown that the states
(\ref{new}) exhibit number squeezing.


\end{document}